\documentclass[sigconf]{acmart}

\usepackage[ruled]{algorithm2e}
\usepackage{stmaryrd}
\usepackage{graphicx}
\usepackage{color}
\usepackage{comment}
\usepackage{url}
\usepackage{wrapfig}
\usepackage{enumerate}
\usepackage{epstopdf}
\begin{document}

\acmYear{2017} 
\setcopyright{acmcopyright}
\acmConference[MMSys'17]{ACM Multimedia Systems 2017}{June 2017}{Taipei, Taiwan}
\acmPrice{15.00}
\acmDOI{http://dx.doi.org/10.1145/3083187.3083211}
\acmISBN{978-1-4503-5002-0/17/06}

\linespread{0.965}

\title{A Mobile Geo-Communication Dataset for Physiology-Aware DASH in Rural Ambulance Transport}

\author{Mohammad Hosseini}
\affiliation{\institution{UIUC}}
\email{shossen2@illinois.edu}

\author{Yu Jiang}
\affiliation{\institution{Tsinghua University}}
\email{jy1989@mail.tsinghua.edu.cn}

\author{Ali Yekkehkhany}
\affiliation{\institution{UIUC}}
\email{yekkehk2@illinois.edu}

\author{Richard R. Berlin}
\affiliation{\institution{Carle Foundation Hospital}}
\email{richard.berlin@carle.com}

\author{Lui Sha}
\affiliation{\institution{UIUC}}
\email{lrs@illinois.edu}

\begin{abstract}
Use of telecommunication technologies for remote, continuous monitoring of patients can enhance effectiveness of emergency ambulance care during transport from rural areas to a regional center hospital. However, the communication along the various routes in rural areas may have wide bandwidth ranges from 2G to 4G; some regions may have only lower satellite bandwidth available. Bandwidth fluctuation together with real-time communication of various clinical multimedia pose a major challenge during rural patient ambulance transport.

The availability of a pre-transport route-dependent communication bandwidth database is an important resource in remote monitoring and clinical multimedia transmission in rural ambulance transport. Here, we present a geo-communication dataset from extensive profiling of 4 major US mobile carriers in Illinois, from the rural location of Hoopeston to the central referral hospital center at Urbana. In collaboration with Carle Foundation Hospital, we developed a profiler, and collected various geographical and communication traces for realistic emergency rural ambulance transport scenarios. Our dataset is to support our ongoing work of proposing ``physiology-aware DASH", which is particularly useful for adaptive remote monitoring of critically ill patients in emergency rural ambulance transport. It provides insights on ensuring higher Quality of Service (QoS) for most critical clinical multimedia in response to changes in patients' physiological states and bandwidth conditions. Our dataset is available online \footnote{http://web.engr.illinois.edu/\textasciitilde shossen2/carleProfiler.html} for research community.
\end{abstract}


\maketitle

\section{Introduction}
\begin{figure}[!b]
\centering
\includegraphics[width=.9\columnwidth]{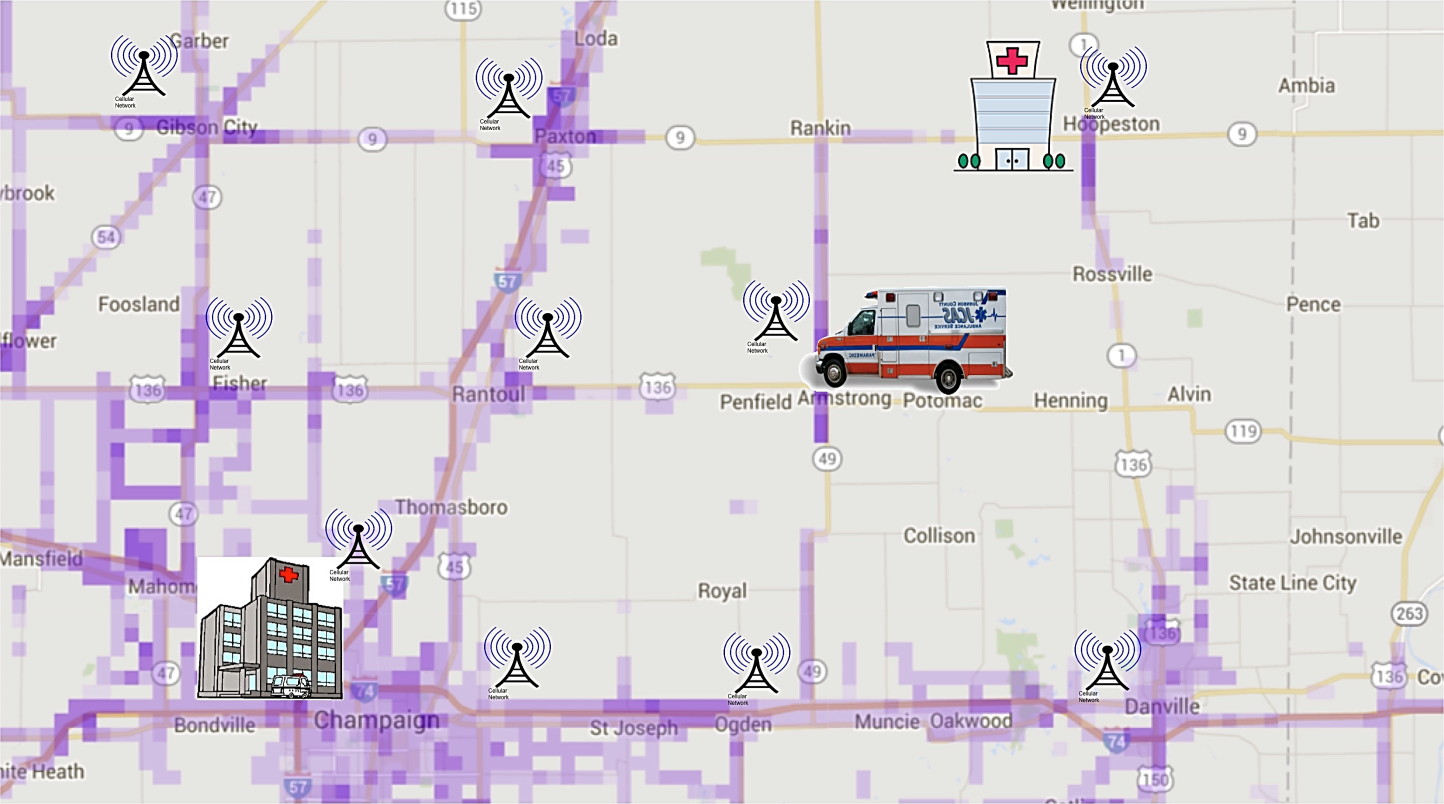}
\caption{Real-world geographical map of our experimental region between Hoopeston rural hospital and Carle center hospital in Illinois.}
\vspace{-.3cm}
\label{fig:coverage}
\end{figure}

\begin{figure*}[!t]
\centering
\includegraphics[width=.75\textwidth]{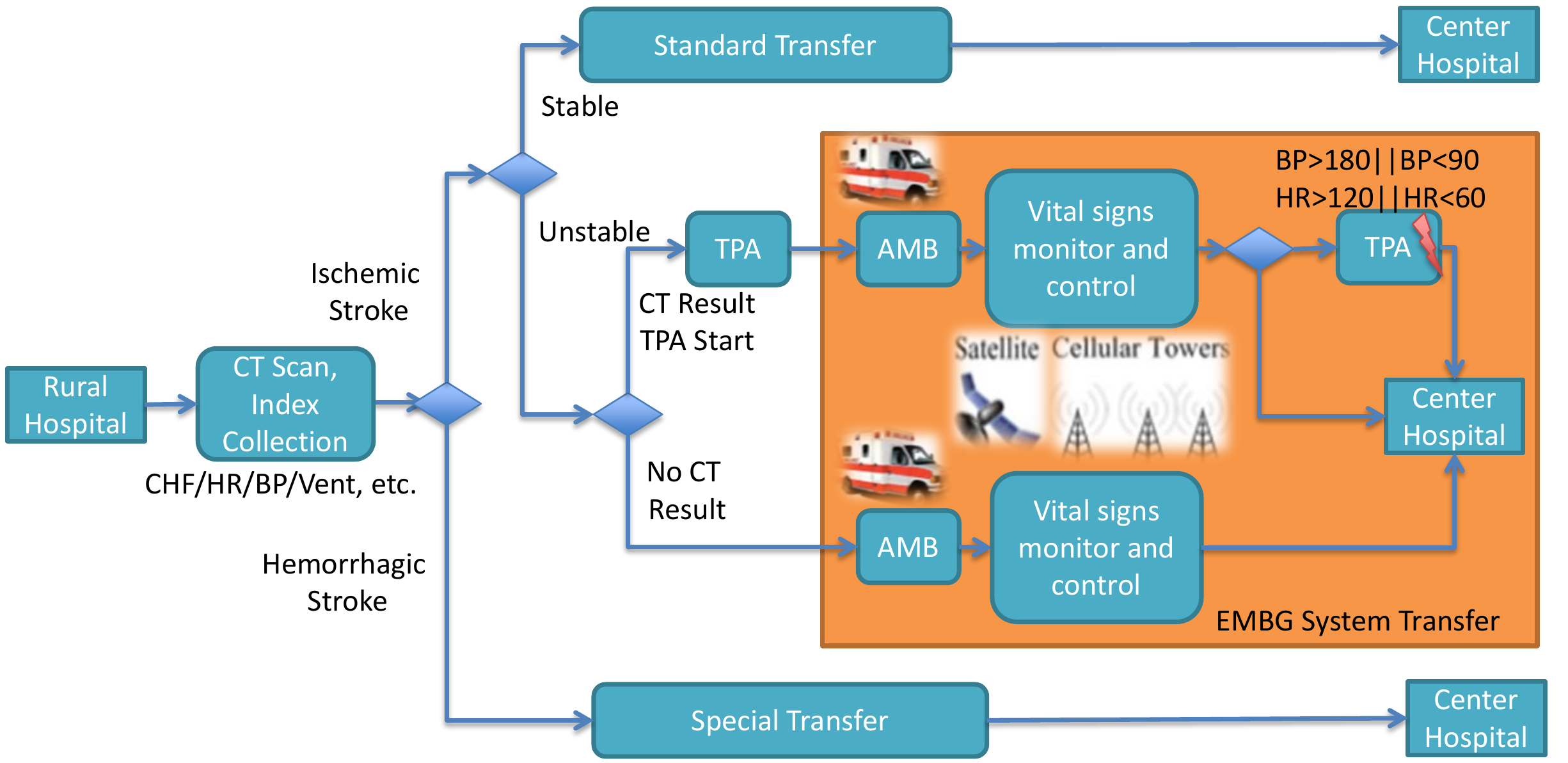}
\caption{Clinical workflow for stroke patient care from a rural hospital to a regional center hospital.}
\vspace{-.3cm}
\label{fig:example-tpa}
\end{figure*}

There is a great divide in emergency medical care between rural and urban areas. The highest death rates are found in rural counties. This together with the higher unavailability of medical tertiary care expertise and pre-hospital services in rural areas have motivated huge research efforts in recent years to enhance the effectiveness of remote patient care.

For mobile care during emergency ambulance patient transport from rural areas to center tertiary hospitals, reliable and real-time communication is essential. It allows the physicians in the center hospital to remotely supervise the patient in the ambulance and assist the Emergency Medical Technicians (EMT) to follow best treatment practices based on patient's clinical states. Unfortunately, remote monitoring of patients involves real-time transmission of various clinical multimedia including clinical videos, medical images, and vital signs, which requires use of mobile network with high-fidelity communication bandwidth. However, the wireless networks along the roads in rural areas range from 4G to 2G to low speed satellite links with high degree of bandwidth fluctuation, which poses a significant challenge to transmit critical clinical multimedia. The problem is exacerbated particularly in the high mobility scenarios of high-speed ambulances. Therefore, ensuring proper QoS for the life-critical and bandwidth-sensitive remote mobile care applications, especially for critical physiological data becomes crucial.

In this paper, we follow our previous work on an adaptive physiological communication architecture \cite{joms2016}, and present our geo-communication dataset that we obtained from extensive profiling of multiple mobile carriers in Illinois. We are mainly motivated by the bandwidth fluctuation and the high QoS requirement of clinical multimedia communication during high-speed ambulance transport as well as the high mortality rate of patients in Illinois's rural environment. In collaboration with Carle Foundation Hospital \cite{carle} in Urbana, we developed a profiler, and profiled various geo-communication information for a realistic emergency ambulance transport targeting a large rural area in Illinois, from Hoopeston's Regional Hospital \cite{hoopeston} to the Carle's center hospital in Urbana. Hoopeston Regional Health Center (rural hospital) is an integrated part of Carle Foundation Hospital in Urbana (center hospital), which includes medical clinic based in Hoopeston, Illinois, with multiple additional clinics serving its surrounding rural communities. 

Figure \ref{fig:coverage} illustrates a real-world geographical map of our experimental region with mobile data coverage map, with darker colors showing higher bandwidth. Our geo-communi-cation dataset includes profiling of 4 major mobile carriers in the US through continuous sampling of communication bandwidths (download and upload rates), GPS coordinates, GPS accuracy, altitude, and vehicle's velocity and bearing. Our dataset is particularly useful to enhance the remote monitoring of patients and the clinical multimedia communication issues during high-speed ambulance transport in large rural hospital settings. Various clinical multimedia can have different priorities depending on the patient's physiological states in context of specific acute diseases. We use our dataset to support our proposition the notion of ``\textit{physiology-aware DASH}" \cite{tmm2017} to extend DASH towards physiology awareness, and study the semantics relation between dynamically changing bandwidth and the physiological priority of communicated clinical multimedia. The insights from our dataset can be crucial in improving the effectiveness of remote patient care through adaptive transmission of various clinical multimedia and transmit more critical clinical multimedia with higher bitrate in response to varying bandwidth as well as changes in the disease and patient's clinical states. Further, transport during critical illness can make route selection patient physiological state dependent; prompt decisions which weigh a longer more secure bandwidth route versus a shorter, more rapid route with less secure bandwidth must be made. Our dataset is particularly useful for research groups intersecting with mobile and wireless communications, multimedia streaming, and tele-medicine sub-divisions. Apart from personal and research purposes, our dataset has real-world applications in ambulance services especially those seen in emergency rural scenarios. We are currently using our dataset to validate our adaptive clinical multimedia transmission system that will serve at central and southern Illinois with 1.2 million people.

Our dataset is unique in the sense that:
\begin{itemize}
\item It specifically targets clinical multimedia communication in emergency ambulance services, with the main goal of enhancing remote patient monitoring through adaptive clinical multimedia transmission.
\item It includes profiling of a large rural environment and rural Illinois in specific with almost 54,000 samples, with a real clinical use-case and hospital collaboration.
\item Prior datasets separately covered measurements of individual geo-communication information, whereas in our dataset we profile a comprehensive and integrated set of all fields as a single dataset.
\item Our communication traces are collected through profiling of 4 different mobile carriers as opposed to only one.
\end{itemize}

To the best of our knowledge, there is no previous geo-communic-ation dataset with all these four features at the same time.

The paper is organized as follows. In Section \ref{background}, we cover some background and related work, and discuss a real clinical use-case in emergency ambulance transport where our dataset can be employed. Section \ref{design} explains the design of our geo-communication profiler and the structure of the measurements. In Section \ref{analysis}, we present a sample of our dataset measurements and discuss some obtained insights, while in Section \ref{conclusion} we conclude the paper.

\section{Background and Related Work}
\label{background}
There has been a large body of work done around capturing datasets targeting various networked services. In \cite{dashdataset1} presented a dataset used for DASH. Their DASH dataset includes metadata for media presentations, providing insights on the advantages as well as problems of various video segment lengths. For a similar application, in \cite{dashdataset2} the authors present a distributed dataset for the DASH standard which is mirrored at different sites within Europe. The dataset was mainly used for simulation of switching between different content delivery networks.

There are also a large body of work using location-based datasets, however, not necessarily generating such datasets per se. These works are mostly using outdoor movements with GPS traces targeting various applications, including sharing of travel experiences \cite{gpsdataset1}, personalized travel recommendations \cite{gpsdataset5}, life logging \cite{gpsdataset3}, user speed estimation \cite{gpsdataset6}, detection of taxi trajectory anomalies \cite{gpsdataset7} and analyzing sports activities \cite{gpsdataset4}.

Probably the most related set of traces to our dataset is \cite{dashdataset3}. In \cite{dashdataset3}, the authors presented measurements of network throughput when adaptive HTTP streaming is performed over 3G networks using mobile devices. They used popular commute routes in Oslo, Norway under different types of public transportation (metro, tram, train, bus and ferry) for profiling of network behavior. Their log provides the GPS coordinates and the number of bytes downloaded for every second in the route.

In our previous work \cite{joms2016}, we proposed an adaptive clinical communication architecture, and designed a physiological message-exchange architecture for emergency patient transport from a rural hospital to a regional center hospital. Our collection of traces in this dataset paper follows our previous work to support and validate the design of our physiological message-exchange architecture. It also supports our ongoing work of proposing ``physiology-aware DASH" \cite{tmm2017} to extend DASH towards prioritized physiology awareness for the purpose of adaptive clinical multimedia transmission with higher bitrate assigned to more critical clinical multimedia given patient's physiological states and available bandwidth.

\subsection{Real-World Clinical Use-Case}
Let's elaborate on emergency care for an acute disease as a real-world use-case and illustrate how geo-communication information especially bandwidth can get crucial within the context of adaptive clinical multimedia communication. To clarify the concepts, we take acute stroke care being practiced at Carle's hospital networks as a real-world example of emergency rural ambulance transport.

Stroke is the third leading cause of death and the first leading cause of disability in the United States~\cite{stroke-stats}. In addition, stroke patients are often elderly (in fact, 65\% to 72\% stroke patients are over age 65~\cite{elderly-stroke-rate}) who may need the highest communication requirement for remote monitoring due to complicating medical factors. Furthermore, some effective stroke treatment medications have strict remote monitoring implementation guidelines; these may begin at the remote facility and continue through ambulance transport to the receiving regional hospital center. Overall, different clinical multimedia may take priority when considering limited communication coverage and bandwidth availability, which therefore must be communicated with higher fidelity.

Figure \ref{fig:example-tpa} illustrates the workflow for a stroke patient being transferred from a rural facility to a regional hospital center. Let's consider a 70 year old male patient arrives at a rural hospital and the diagnosis of acute stroke is considered. A CT (Computerized Tomography) head scan is performed. The CT images are sent electronically to the regional center for interpretation. At this clinical state, transmission of streams associated with CT images has highest priority. The patient's neurological examination, laboratory data, and vital signs including heart rate (HR), blood pressure (BP), oxygenation level are obtained and sent for the purpose of continuous monitoring. The diagnosis of an acute stroke is made, and the patient is placed in an ambulance for transport with physicians at the center hospital remotely monitoring the patient.
\begin{figure}[!t]
\centering
\includegraphics[width=.9\columnwidth]{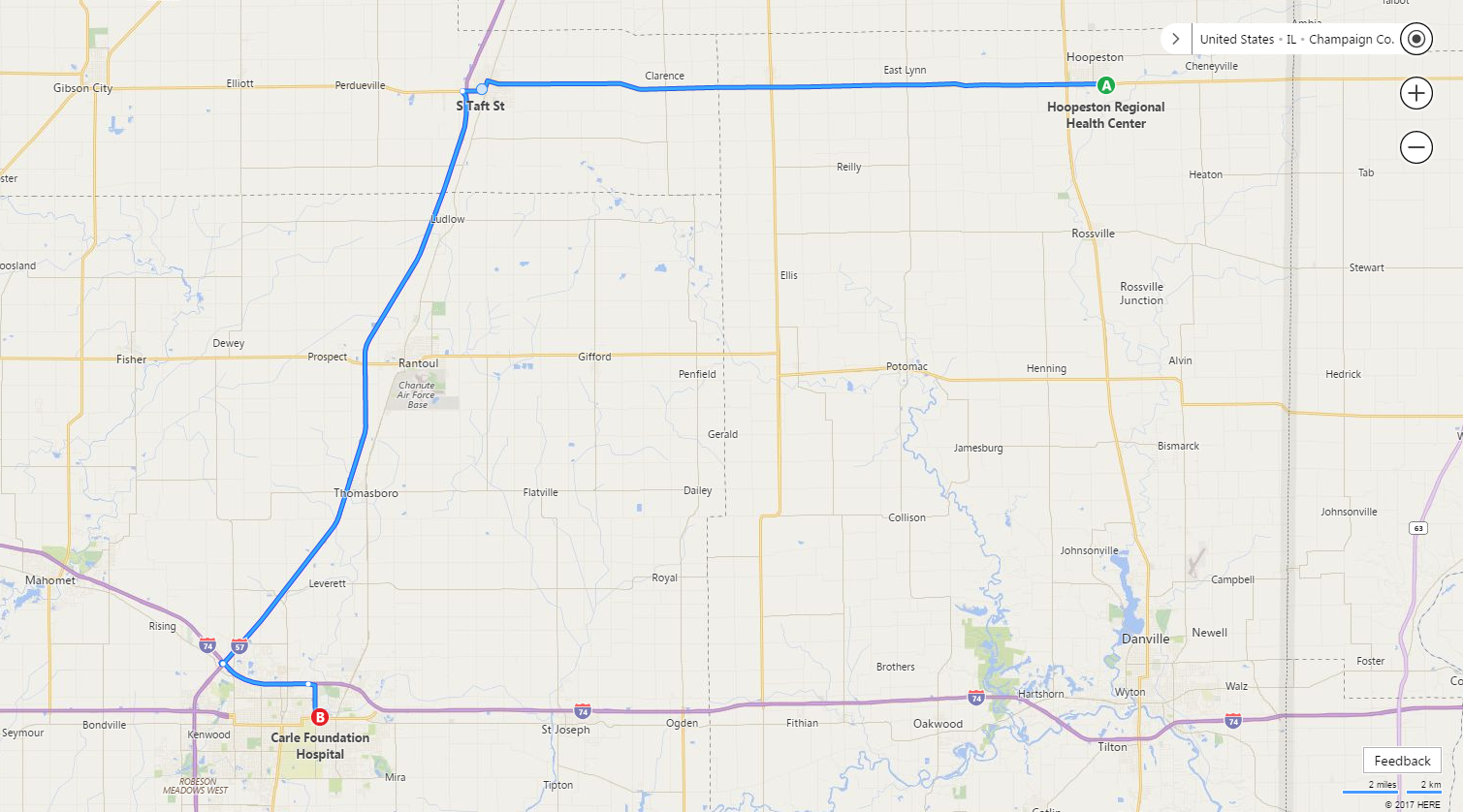}
~\\
~\\
\includegraphics[width=.9\columnwidth]{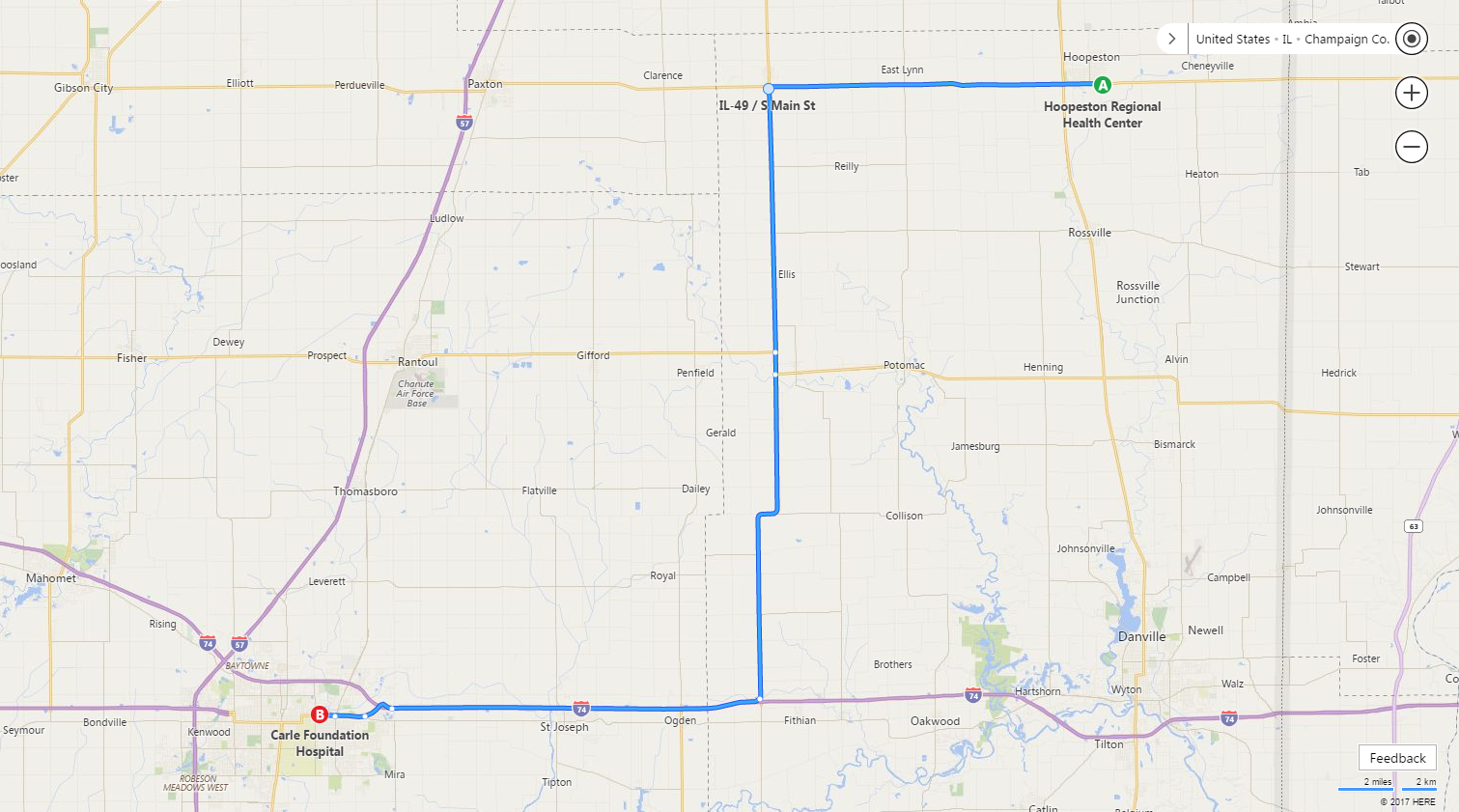}
\caption{Abstract road maps of our profiled region.}
\vspace{-.5cm}
\label{map}
\end{figure}
A video camera and microphone in the ambulance, connected to the regional center, is used to remotely monitor the patient's physiological status en route. It is determined that the patient has a hemorrhagic stroke (bleeding into the brain from blood vessel rupture). In this case, clinical best practices suggest that time of transport gets most important. Available bandwidth is used for communication with specialists at the regional center in case emergency consultation, interventional radiology or surgery is indicated. The HR, BP, oxygenation, and neurological status are remotely monitored. In these situations, with higher bandwidth, the audio/video support and therefore, the audio-visual streams take higher bitrate. However, in the event of limited bandwidth, the audio-visual monitoring system as well as the transmission of repeated laboratory data gets limited with secondary priority. Highest priority in this situation is maintaining the patient's vital signs. The HR and BP in specific must be kept within strict limits. The BP assumes particular importance if it rises too high (greater than 180/-) or falls too low (less than 90/-) which are indicated by the physiological models. In accordance with guidance given by best-practice received, audio communication with the center hospital to manage elevated blood pressure assumes highest priority if BP is greater than 180. This might require the continuous intravenous infusion of active medications to lower blood pressure in the ambulance, using nicardipine or nitroprusside medications where vigorous real-time monitoring is needed. Once higher communication bandwidth is available, periodic laboratory data, treatments for an elevated blood glucose (greater than 350), and the video camera streams can be transmitted with normal transmission frequency. Overall, a continuous and high-fidelity communication coverage gets crucial depending on how critical remote monitoring can be. Use of geo-communication information therefore can provide extremely useful insights on ensuring higher QoS for remote monitoring through either adapting the clinical multimedia transmission bitrate depending on the priority of the clinical data, or even selecting best routes for ambulance transport or adjusting ambulance velocity in response to changes in patients' physiological states and bandwidth conditions.

\begin{wrapfigure}{R}{0.4\columnwidth}
\setlength{\columnsep}{.5cm}
  \centering
  \vspace{-.3cm}
  \hspace{-.5cm}  \includegraphics[width=.45\columnwidth]{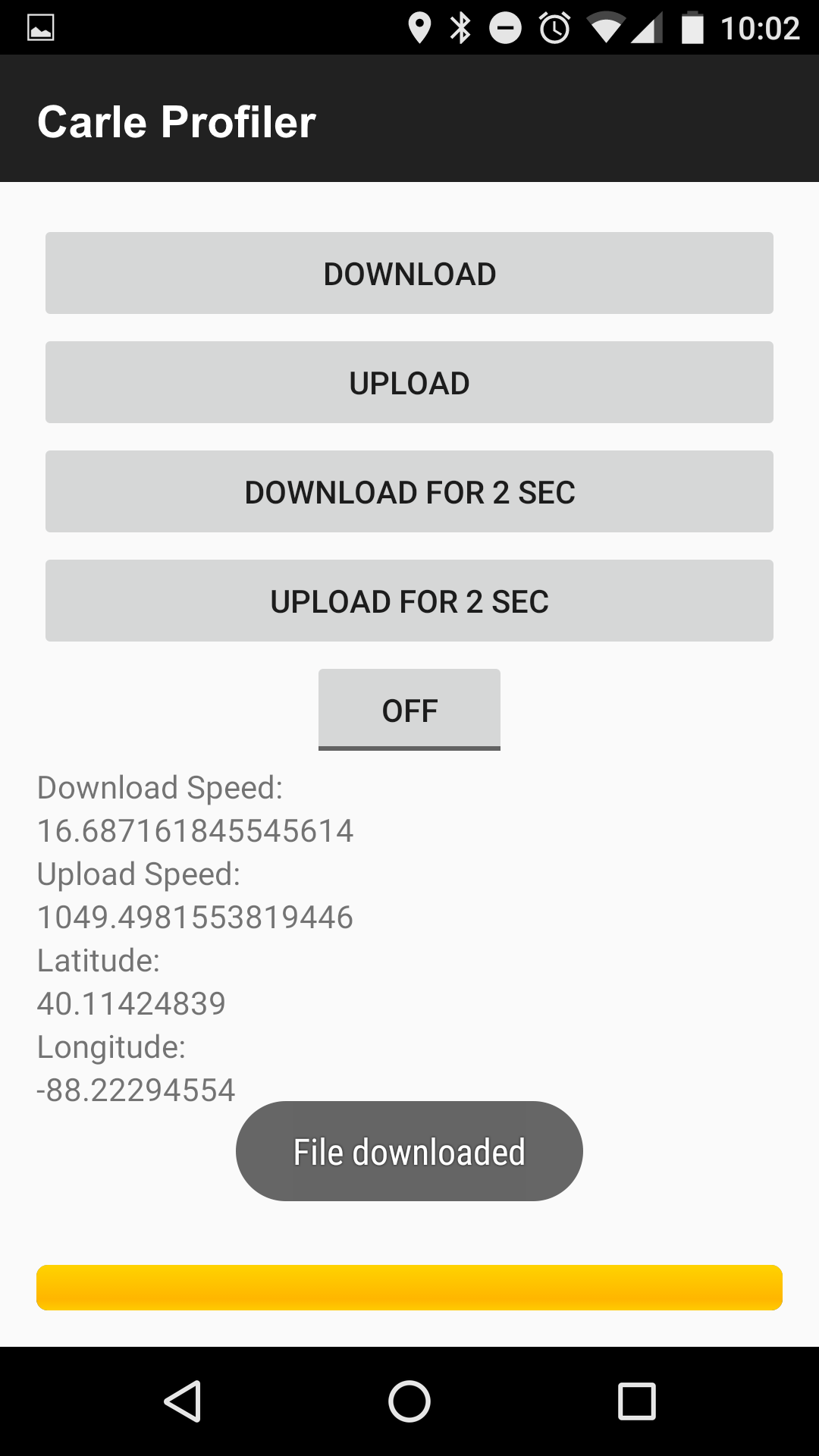}
    \caption{Our developed profiler.}
    \label{profiler}
      \vspace{-.5cm}
\end{wrapfigure}

\section{System Design and Structure}
\label{design}
To collect our geo-communication traces, we conducted our profiling experiments for major routes within rural hospital settings to assess the effectiveness and to facilitate analyzing a logical interpretation of how our physiological communication architecture works in real world. We collected almost 54,000 samples in 4 trips under 4 major US cellular carriers, by driving through two major routes covering a large rural hospital environment in Illinois, from Hoopeston's rural hospital to Carle's center hospital in Urbana. Figure \ref{map} illustrates the geographical trajectories of the two profiled routes.

\subsection{Profiler Development}
To collect our geo-communication information, we developed a mobile geo-communication profiler application in collaboration with Carle Ambulance Service \cite{carle}. We used Android SDK 25 for development, and used Google Nexus 5 and Google Nexus 5X smartphones as our profiling platform mounted with 4G LTE ICCID SIM Cards under 4 major cellular carriers in the US: Sprint, AT\& T, T-Mobile, and Verizon.

\begin{table*}[!htb]
\centering
\caption{An example set of 5 consecutive geo-communication traces (Sprint)}
\label{table:sample}
\begin{tabular}{|l|l|l|l|l|l|l|l|l|}
\hline
\multicolumn{1}{|c|}{{\bf Timestamp}} & \multicolumn{1}{|c|}{{\bf Downlink}} & \multicolumn{1}{|c|}{{\bf Uplink}} & \multicolumn{1}{|c|}{{\bf Longitude}} & \multicolumn{1}{|c|}{{\bf Latitude}} & \multicolumn{1}{|c|}{{\bf Accuracy}} & \multicolumn{1}{|c|}{{\bf Altitude}} & \multicolumn{1}{|c|}{{\bf Velocity}} & \multicolumn{1}{|c|}{{\bf Bearing}}                                    
\\ \hline
100 & 1192.1749 & 971.52216 &  40.459123 & -88.070278 & 14 & 193.0 & 28.75 & 91.1
\\ \hline
101 & 566.8956	& 932.59045 &   40.459017 & -88.066069 & 13 & 189.0 & 30.0 & 90.7
\\ \hline
102 & 834.95325	& 1560.9518 &  40.459028 & -88.063833 & 11 & 195.0 & 30.75 & 90.6
\\ \hline
103 & 1192.1749 & 971.52216 &  40.459123 & -88.070278 & 14 & 193.0 & 28.75 & 91.1
\\ \hline
104 & 878.0356	& 1117.8868 &  40.459036 & -88.062363 & 11 & 197.0 & 30.75 & 90.3
\\ \hline
\end{tabular}
\end{table*}

Figure \ref{profiler} shows a screenshot of our developed geo-commun-ication profiler. Our profiler periodically samples and logs various useful geo-communication information once every 4 seconds (2 seconds dedicated for download rates and 2 seconds for upload rates). The profiled geo-communication information includes: a) timestamp, b) downlink bandwidth, c) uplink bandwidth, d) GPS longitude, e) GPS latitude, f) GPS accuracy, g) altitude, h) velocity, and i) bearing (the bearing from the source to the destination location in degrees east of the true north). All traces were stored locally on the profiling device and were collected at the end of each experiments. Table \ref{table:sample} shows 5 consecutive samples of our dataset, containing values for each of these fields respectively.
\begin{figure}[!t]
\centering
\hspace{-.2in}\includegraphics[width=1.\columnwidth, trim = 50 118 60 120, clip = true]{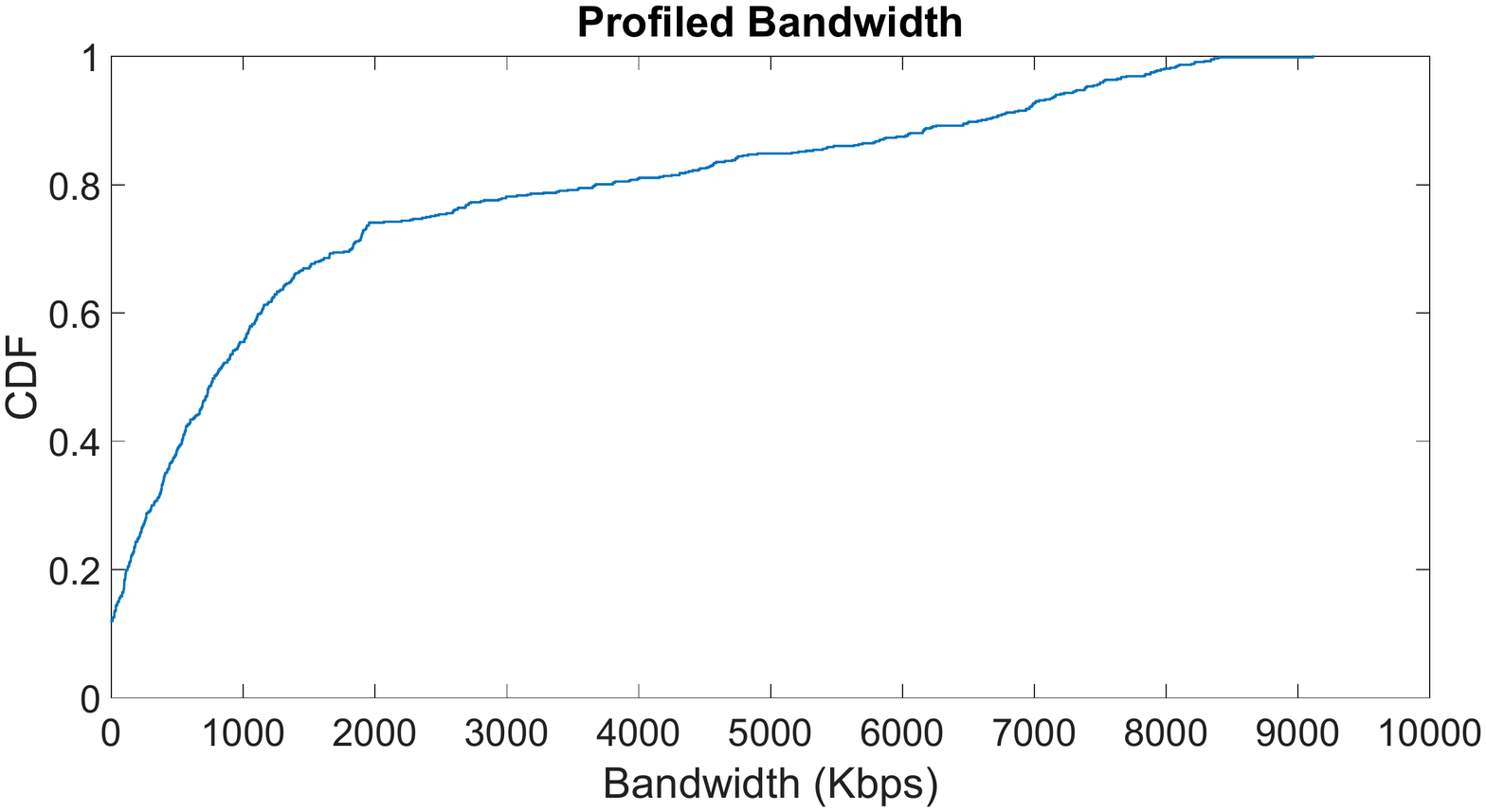}
~\\~\\\vspace{-.2cm}
\hspace{-.2in}\includegraphics[width=1.\columnwidth, trim = 50 118 60 120, clip = true]{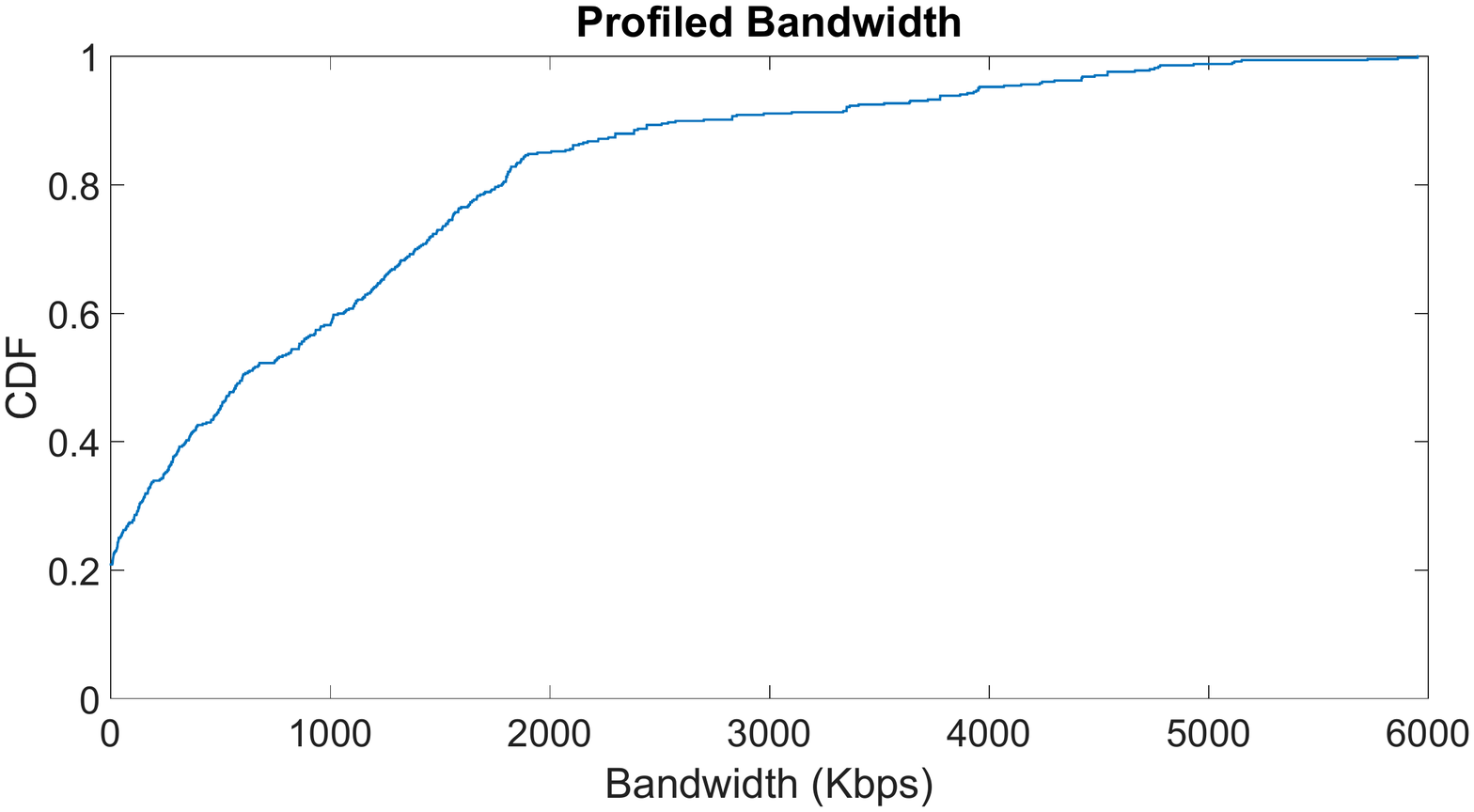}
\vspace{-.3cm}
\caption{CDF of bandwidth traces for two samples.}
\vspace{-.5cm}
\label{cdf}
\end{figure}
To implement the profiling process for downlink rates, we followed a similar approach to Ookla's SpeetTest \cite{ookla}. our profiler client first establishes multiple TCP connections with our server over port 8080, and continuously retrieves file chunks of 1 megabits (Mb) for a course of 2 seconds. Our server is a HTTP server that we specified to minimize latency and jitter due to congestion and communication errors. As the chunks are received by the profiler, the profiler requests more file chunks throughout the fixed duration. The total size of the buffered transfers is then received, and the download rate is calculated in kbps (1 byte = 0.0078125 kilobits) given the fixed specified duration. The sampling process ends once the configured amount of duration has been reached. For uplink measurements, our profiling works in a similar way. The profiler client first establishes multiple TCP connections with our local server over port 8080, and continuously sends chunks of random generated data in uniform sizes for a course of 2 seconds. The data are then pushed via POST method to the server-side PHP script that we have developed and placed on our server. As the chunks are received by the server, the profiler sends more file chunks throughout the fixed duration. The profiler then retrieves the total size of the buffered transfers, and the uplink rate is calculated given the fixed duration. Similar to the downlink process, the uplink sampling process ends once the configured amount of duration has been reached.\vspace{-.2cm}

\begin{figure*}[!t]
\centering
\includegraphics[width=1.02\columnwidth, trim = 55 250 55 265, clip = true]{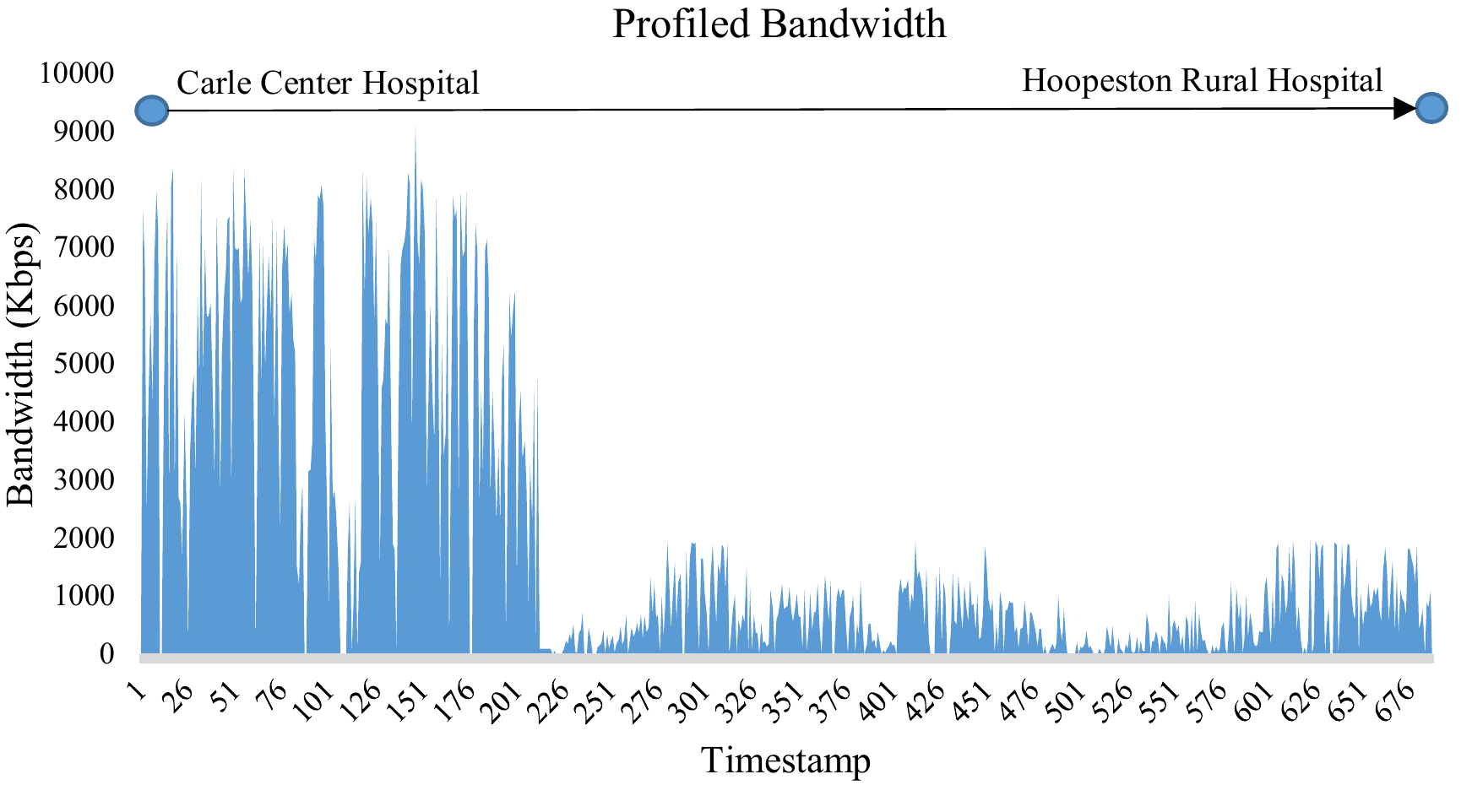}\includegraphics[width=1.02\columnwidth, trim = 55 250 55 265, clip = true]{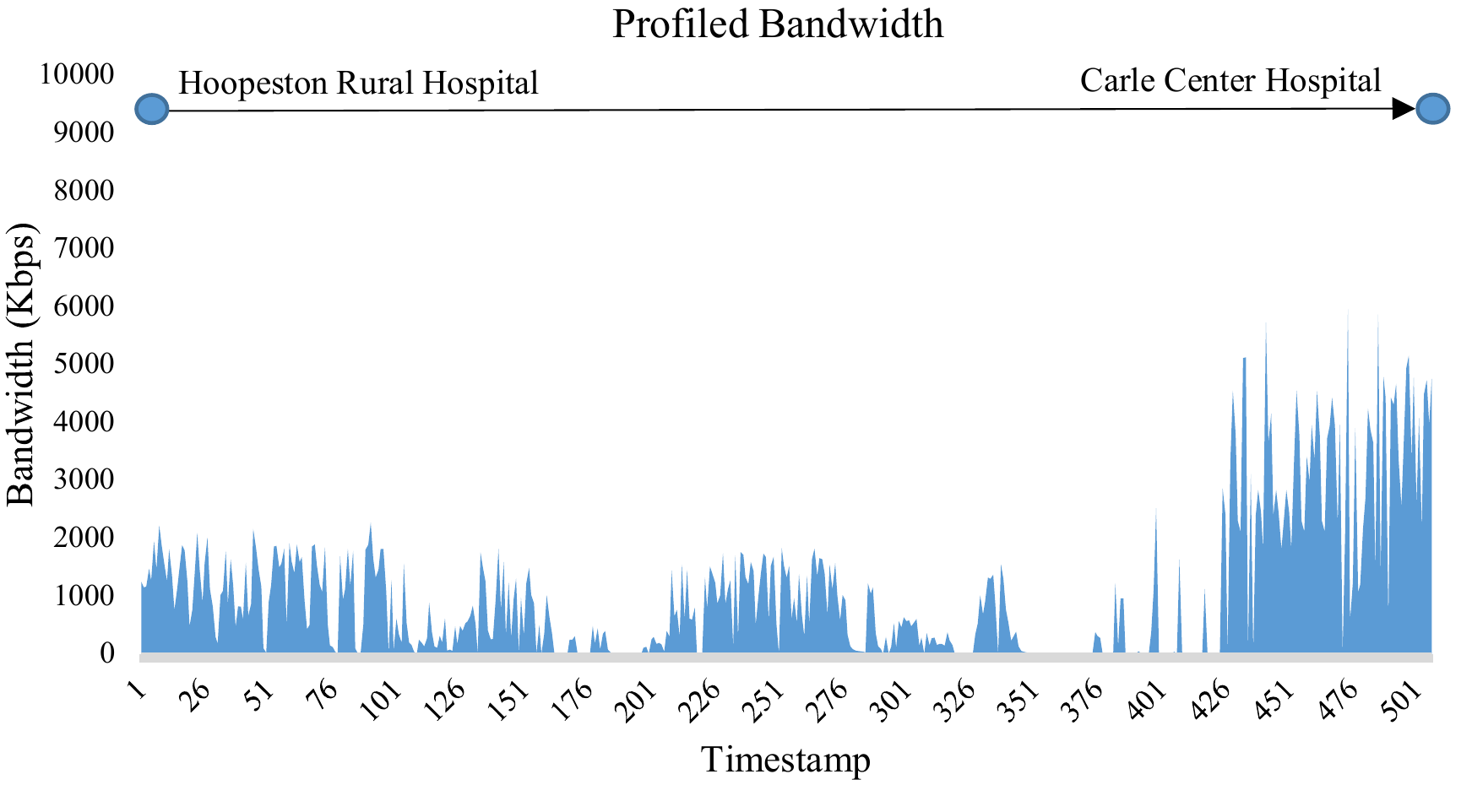}
\includegraphics[width=1.02\columnwidth, trim = 55 250 55 265, clip = true]{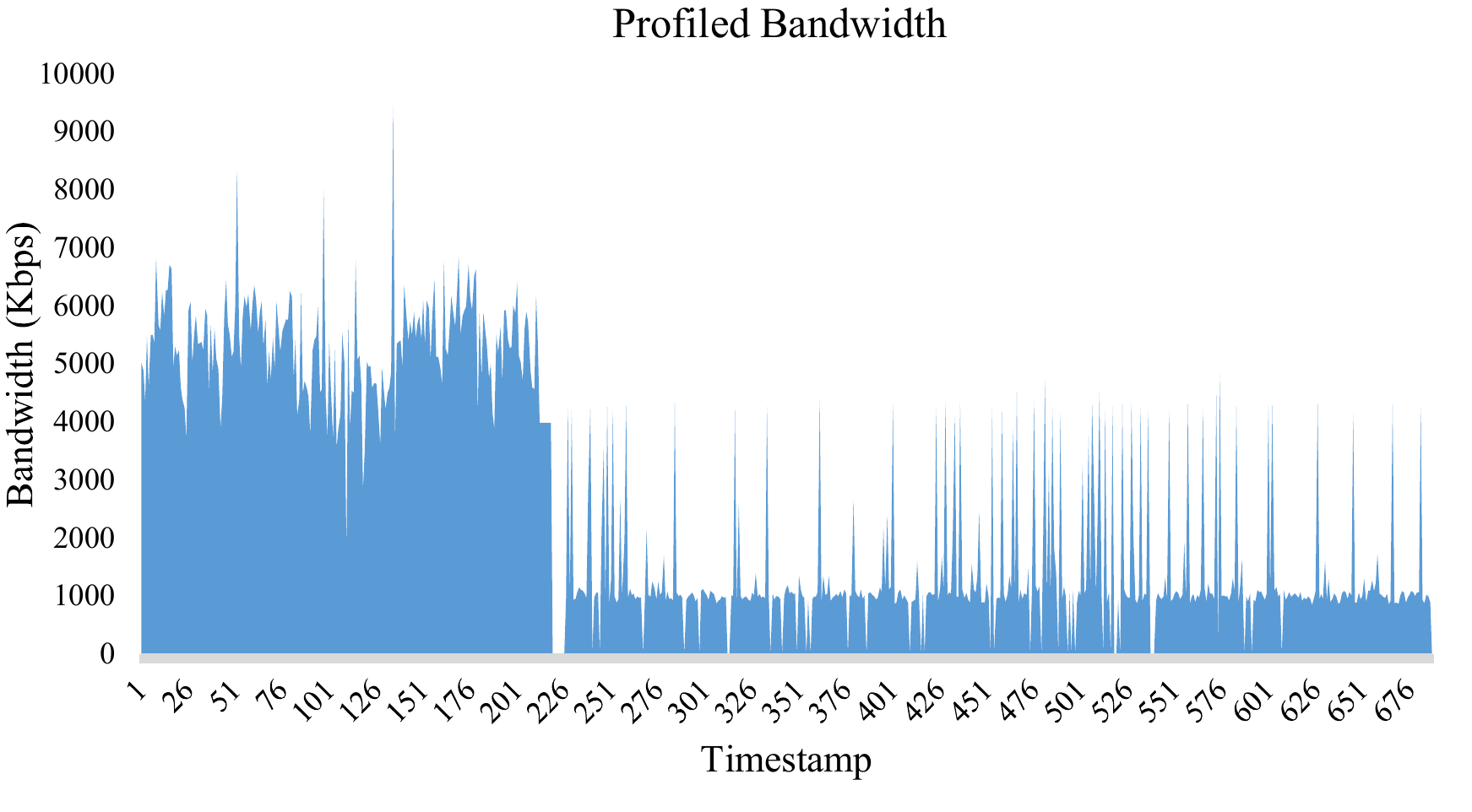}\includegraphics[width=1.02\columnwidth, trim = 55 250 55 265, clip = true]{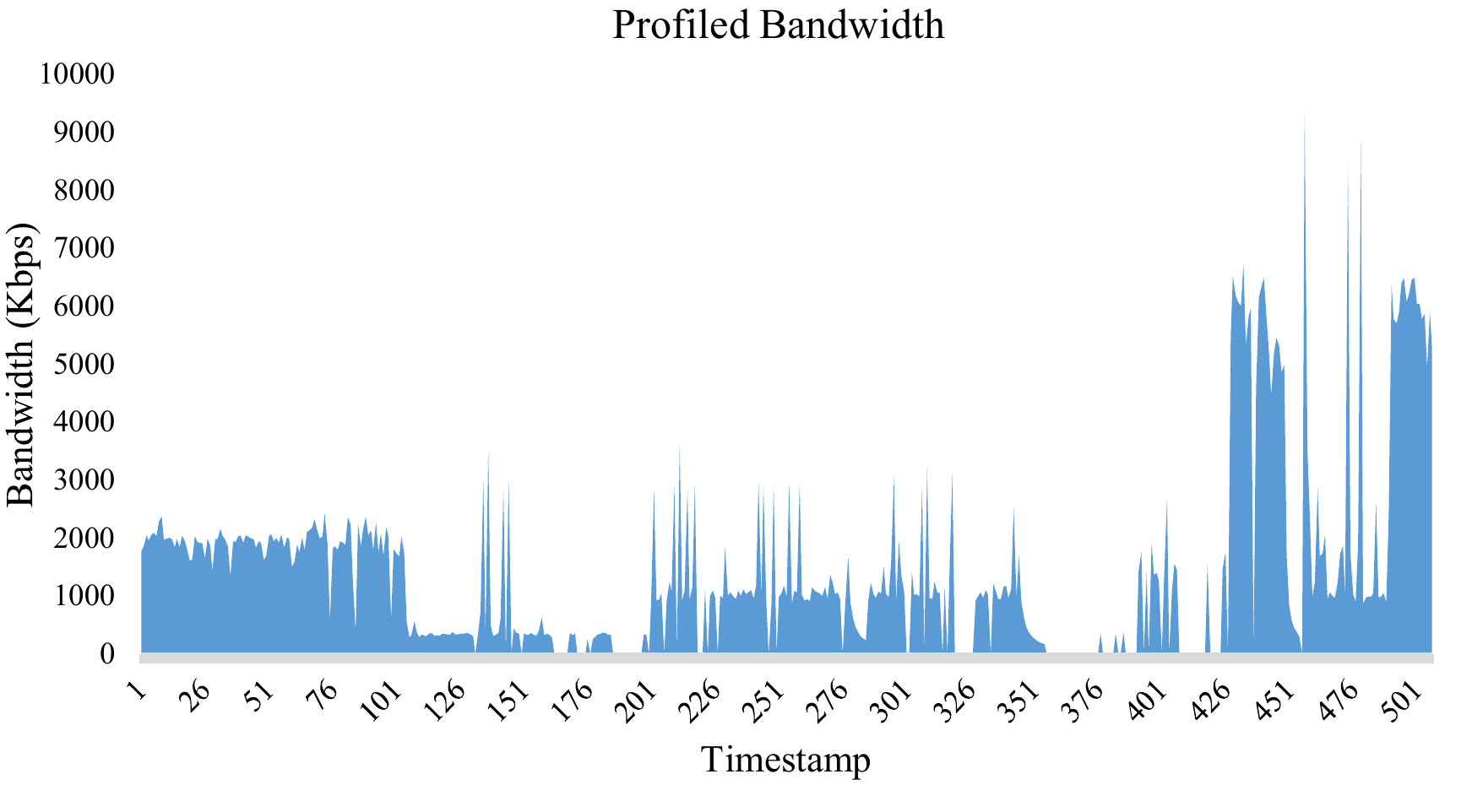}
\includegraphics[width=1.02\columnwidth, trim = 65 250 55 265, clip = true]{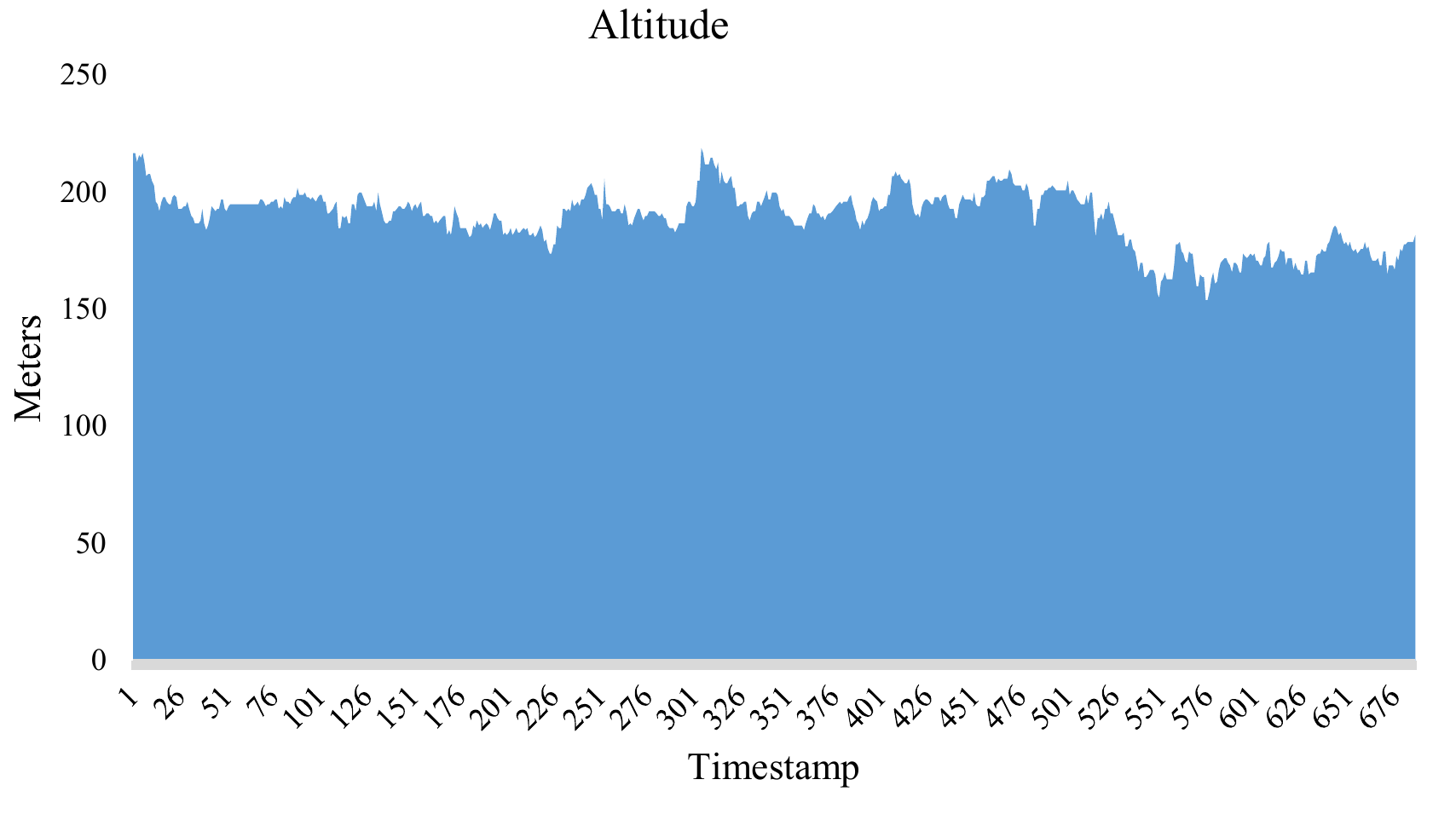}\includegraphics[width=1.02\columnwidth, trim = 70 250 55 265, clip = true]{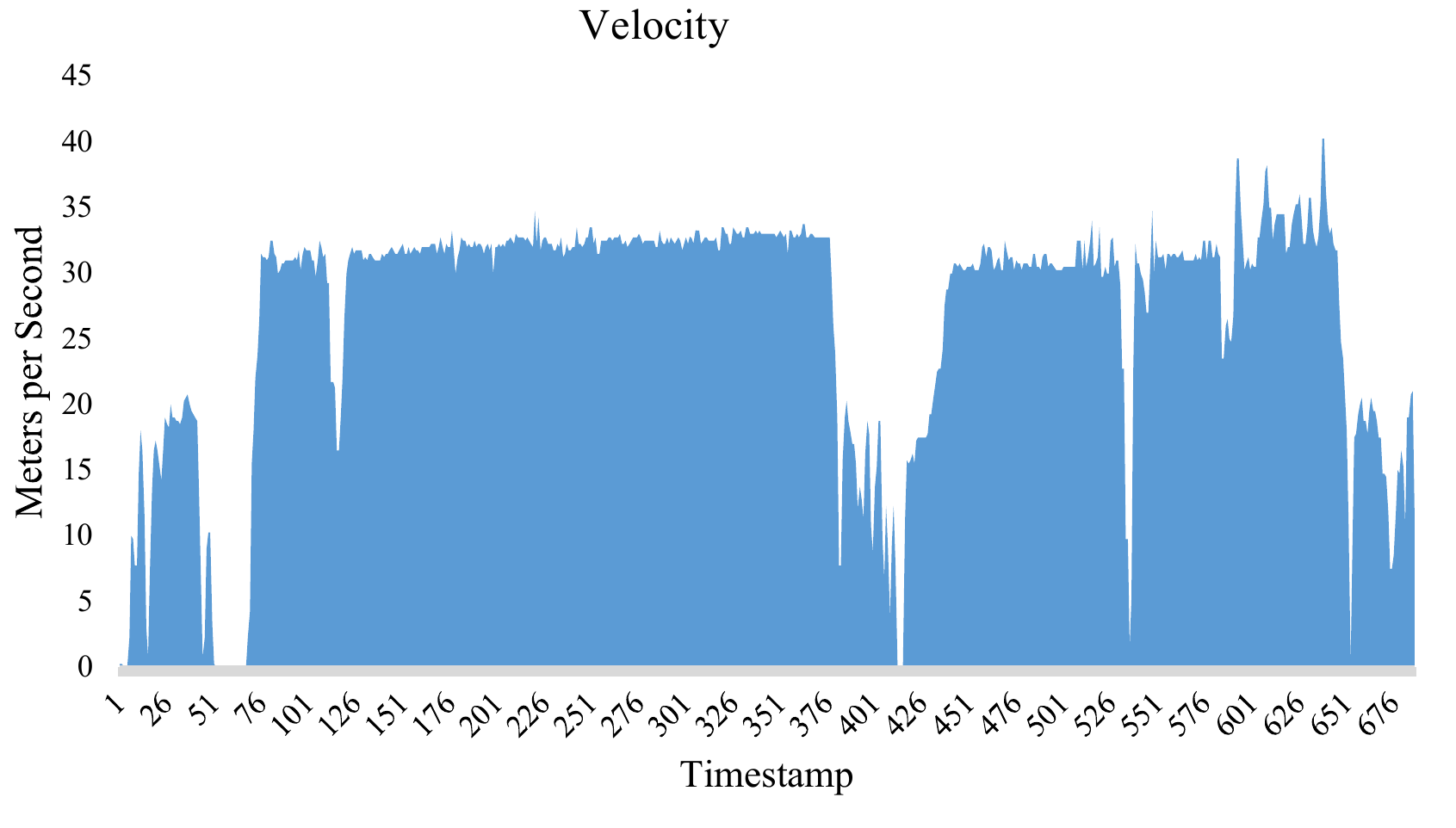}
\vspace{-.3cm}
\caption{A sample of our collected geo-communication traces from Hoopeston rural hospital to Carle center hospital.}

\label{profile}
\end{figure*}

\section{Analysis of Traces}
\label{analysis}
Figures \ref{cdf} and \ref{profile} demonstrate only a small subset of our dataset under Sprint cellular network, 
presenting a subset of all our profiled communication and geographical information. Figure \ref{profile} (Top-Left) and (Middle-Left) show the downlink and uplink bandwidths for the longer route shown in Figure \ref{map} (Top). For the purpose of comparisons, Figure \ref{profile} (Top-Right) and (Middle-Right) show the downlink and uplink bandwidth for the shorter route shown in Figure \ref{map} (Bottom). The vertical axis in all these four figures shows the available bandwidth in Kbps while the horizontal axis shows the timestamp with each point of data accounting for the four seconds of sampling period. The figures showcase interesting results to support our assumption of bandwidth \textit{variability} and \textit{limitation}. As can be clearly observed in all four figures, overall the results show lower communication bandwidth on rural areas while they show higher bandwidth as we get closer to the urban areas. It can clearly be seen that the communication bandwidth along both routes can range from as low as a few Kbps to a few Mbps, with most part of the routes suffering from very poor communication coverage, and some parts, especially in the shorter route, with no communication coverage. In Figure \ref{cdf}, we present the empirical CDFs of downlink bandwidth in both routes to better illustrate the concept of limited bandwidth. As can also be seen here, more than \textit{half} of the both routes suffer bandwidth rates of less than 1000 Kbps, with almost \%17-\%23 of the traces showing no data coverage (0 Kbps). The low communication bandwidth severely limits the remote monitoring capability and therefore, transmission of clinical multimedia that can be communicated during emergency ambulance patient transport in rural areas. As can be witnessed, interestingly while the route shown in Figure \ref{map} (Bottom) is shorter with less transport duration, it involves more vigorous communication breakage, making it more suitable for emergency scenarios where transport duration is of higher criticality than the remote monitoring. On the contrary, the longer route shown in Figure \ref{map} (Top) shows a higher fidelity and more continuous communication coverage in spite of longer transport duration, which makes it more suitable for emergency ambulance transports where remote monitoring becomes more critical. This particular information can provide useful insights on designing a disease-aware scheduler for next-generation ambulance dispatch centers which can assist ambulances pick the best route when continuous network coverage is critically needed for real-time monitoring of patients. 

Figure \ref{profile} (Bottom-Left) illustrates the altitude distribution in meters for the route shown in Figure \ref{map} (Top). It shows the variance in altitude in our experimental regions is low (standard deviation of $\%$5.8), interestingly proving the fact that Illinois is mostly flat prairie and hill-less plains, in fact the second flattest state on mainland \cite{illinoisisflat}. Figure \ref{profile} (Bottom-Right) illustrates the vehicle's velocity traces during that specific instance of profiling. The average travel speed during the profiling process was 27.93 meters per second (62.5 miles per hour). It is expected that the higher travel speed of an ambulance during an emergency situation can further limit the available communication bandwidth. 
\section{Conclusion and Discussion}
\label{conclusion}
Use of telecommunication technologies for remote monitoring of patients can enhance effectiveness and safety of emergency ambulance transport from rural areas to a regional center hospital. However, the communication along the roads in rural areas can be as low as a few Kbps, with some parts with no communication coverage. This bandwidth fluctuations together with the real-time communication of various clinical multimedia can pose a major challenge in remote supervision of patients. Use of geographical and communication statistics can therefore assist the EMT to associate best treatments through a better prediction of the communication behavior and help them take proper actions depending on the patients physiological states.

In this paper, we present a geo-communication dataset from extensive profiling of multiple US mobile carriers in a large rural area in Illinois, from Hoopeston to Urbana. In collaboration with Carle Ambulance Service, we developed a profiler, and collected almost 54,000 samples of various geographical and communication traces targeting a realistic emergency rural ambulance transport. Our dataset is particularly useful to support remote monitoring of patients in large rural hospital settings. It further provides insights on ensuring higher QoS for remote monitoring of patients through adaptively assigning higher bitrate for most critical clinical multimedia data depending on the physiological state, or even adaptively selecting best routes or adjusting ambulance velocity in trade for a better communication coverage.

Our dataset is available online for research community, and can be useful for various research communities. We believe our traces have high potentials in improving the effectiveness of emergency patient care in major rural hospital settings, urban and suburban, as well as military settings, as it was useful for our real-world use-case in Carle's Ambulance Service in Illinois region. We are currently using our dataset to support our ongoing work of extending DASH towards physiology awareness by validating our adaptive clinical data transmission system that will serve at central and southern Illinois with 1.2 million people.
\section*{Acknowledgments}
This research is supported in part by NSF CNS 1329886, NSF CNS 1545002, and ONR N00014-14-1-0717.

\bibliographystyle{ACM-Reference-Format}
\bibliography{sigproc}

\end{document}